# Optical Tests of a 3.7-m diameter Liquid Mirror: Behavior under External Perturbations

Grégoire Tremblay, and Ermanno F. Borra,
Centre d'Optique, Photonique et Laser,
Département de Physique, Université Laval, Québec, Qc, G1K 7P4

email : borra@phy.ulaval.ca, tremblay@phy.ulaval.ca

(SEND ALL EDITORIAL CORRESPONDENCE TO : E.F. BORRA)

RECEIVED______________________________________

## Abstract

We have built and tested a 3.7-m diameter liquid mirror that rotates on a ball bearing. We have carried out extensive optical tests. We find that although the ball bearing has a poor quality, the quality of the mirror, with mercury layers 1-mm thick, is surprisingly good. Taken at face values the instantaneous Strehl ratios indicate a mirror, that is not quite diffraction limited but usable for astronomical applications. However, the large coning error of the bearing (1.5 arcseconds P-V) induces an excessive wobble, considerably worsening the time averaged PSF. The most interesting result of the interferometry is that we do not see any evidence of the strong astigmatism that may have been expected from Coriolis forces.

We have made good use of the poor quality of the ball bearing to study the effects of vibrations on the surface of the mirror. We have developed a model that reproduces reasonably well the vibration-induced concentric rings seen on LMs. Our studies of wind-induced spirals defects detect the onset of turbulence at smaller Reynold's numbers than we expected. We do not detect neither the spiral defects nor the effects of turbulence for thin mercury layers (<1-mm thick); they dampen them out of detectability.

This work illustrates, once more the crucial importance of working with mercury layers as thin as possible. Large liquid mirrors would have unacceptable optical qualities for mercury layer much thicker than 2 mm.

On the basis of our tests of the 3.7-m mirror it appears that the upper limits to the diameters of LMs having good bearings are above 4 meters. We can say this for the two "show-stoppers" of liquid mirrors, Coriolis forces and self-induced winds, are "no-shows".

**1. Introduction**

Following the suggestion [1] that modern technology renders liquid mirrors useful to astronomy, a feasibility study was undertaken to determine whether, in practice, it is possible to generate an optical quality surface on a spinning liquid. This led to optical tests that showed that a 1.5-m diameter liquid mirror (LM) had such good optical quality that it was diffraction-limited [2].  This article [2] gives a wealth of information on the basic LM technology. It was followed by tests of a 2.5-m diameter liquid mirror [3] that also showed diffraction limited performance. Milestones have been reached by Content et al. [4] who published the first scientific paper to report astronomical research with a liquid mirror telescope (LMT), Hickson et al. [5] who built and demonstrated a professional-quality LMT, and by Cabanac, Borra and Beauchemin [6] who published astronomical research based on a large number of nights obtained with the NASA NODO 3-m [7]. This article [6] is particularly noteworthy for it shows (e.g. their figure 2) that LMTs are sufficiently robust to work for extended periods of time.

Liquid mirrors are interesting in other areas of science besides astronomy. For example, the University of Western Ontario has built a Lidar facility that houses a 2.65-m diameter liquid mirror as receiver [8]. That facility has yielded published scientific research [9]. A Lidar facility has also been built and operated by the University of California at Los Angeles [10]. Liquid mirrors are also used as reference surfaces to test conventional optics [11]. Liquid mirrors have industrial applications. For example, Thibault and Borra [12] have demonstrated a telecentric three-dimensional scanner that uses a liquid mirror. A somewhat outdated review paper [13] gives a convenient summary of the status of LMTs and related issues.

In this article, we discuss the results of extensive tests of a 3.7-m LM. We discuss scattered light measurements and the behavior of the mirror subjected to external perturbations. A more detailed description of the tests can be found in Tremblay's Ph.D. thesis [14]

**2. The mirror**

The basic mirror setup has not changed from the earlier articles [2,3]. Its main components are shown in Fig.2 of [2]. They are a base (a three-point mount) used to align the axis of rotation within a fraction of an arcsecond with respect to the vertical and a bearing on which rests a container that holds the reflective liquid. The system is spun with a synchronous motor driven by an AC power supply. We briefly describe below the mechanical system of the 3.7-m diameter liquid mirror.

A. Three-point mount.

The three-point mount is a sturdy welded-steel triangle designed with finite elements software. The tilt angle of the mount is adjusted by changing the height of two of the points with inexpensive, commercially available, adjustable wedges used to level machine tools. Our deflection measurements show that the mount has the rigidity predicted by the software.

B. Bearings.

We originally planned to use an air bearing. However, because of budgetary limits, we chose an air bearing that had specifications uncomfortably close to our minimum requirements. We were particularly concerned with the ultimate load capacity in tilt of the bearing. During the startup phase, as explained in [2], the mercury surface distribution is noticeably asymmetric, causing an off-centered load that could damage the air bearing. As a matter of prudence, we therefore decided to use a ball bearing during the commissioning of the mirror to practice our mirror-closing skills. During practice, it became obvious that the air bearing would not stand the off-centered load generated by the asymmetries that we saw. We therefore had to give up the air bearing, consequently all of the tests reported in this article were carried out with the ball bearing. On the one hand, this is unfortunate since the quality of the ball bearing is not as good as that of the air bearings that have been previously used, so that we could not obtain the same sort of "clean" results reported in [2] and [3]. On the other hand, using the ball bearing was useful for its poor quality gave us considerable information on the behavior of the mirror under perturbation. It is surprising, and a tribute to its robustness, that the mirror cum ball bearing is almost good enough for astronomical research, as we shall see. We have also briefly experimented with an oil bearing that turned out to have an inadequate design. Since we now have experience with three types of bearings (ball bearings, air-lubricated bearing and oil-lubricated bearings) we summarize our findings below.

B1. Air bearings.

We originally bought a commercial air bearing capable of an ultimate load of 1090 Kg, adequate for our 270 Kg container loaded with the 340 Kg of mercury corresponding to the 2.3-mm thick layer of mercury used at startup. The layer was pumped down to thinner layers in most experiments. Likewise the tilt safety factor of 9.45 (see [13, 15] for a definition of the tilt safety factor) is adequate. However, we found that the off-centered loads generated during startup exceeded the ultimate moment load capacity specified by the builder of the bearing. With our standard setup we protect the air bearing by placing the container, unbolted, on a plate having a diameter such that too large an off-load separates the mirror from the bearing. Metal safety posts placed under the rim hold safety wheels, positioned with a very small gap under the rim, that prevent a catastrophic spill by nudging the mirror back to its position on the bearing. However, in our case, we find that the diameter of the plate would have to be 8 cm, uncomfortably small for a 3.7-m diameter mirror. Paradoxically, the situation may be improved by using a less rigid mount, which flexes and allows the rim to hit the safety wheels during the off-centered load events. This tale gives an important lesson for anyone who wants to build a liquid mirror cum air bearing: The maximum load capacity of the air bearing is a necessary but not a sufficient selection criterion. In this respect, one has to be aware that the ultimate moment load capacity listed in the catalogs is usually for an unloaded bearing and that the ultimate moment load capacity decreases with load. Caveat Emptor!

Earlier tests of liquid mirrors that used air bearings [2,3] did not carry out engineering tests as detailed as in this article. They left several questions unanswered, in particular one regarding whether air bearings dampen vibrations. To answer this question, we carried out vibration tests on a 1.4-m diameter liquid mirror that has been used to build a telecentric 3-D scanner [16]. This mirror has the same air bearing used to test the 2.5-m LMs [3]. We find that the air bearing does not dampen any vibration and does not introduce any vibration: it is neutral. We measured the torque due to friction, obtaining 0.005 Nm.

B2. Ball bearing

Ball bearings have attractive characteristics for they are robust, are relatively inexpensive and need little maintenance. The ball bearing used is actually made of two bearings. The main bearing, is an angular contact thrust ball bearing, double acting (model 234424 BM SP), made by SKF. This bearing only carries the axial load of the mirror and has little angular rigidity. Angular rigidity is attained by adding a second bearing at the end of a 13.95-cm diameter 28.825-cm long solid steel rod that connects the two bearings. This second bearing is a cylindrical roller bearing, double row model NN3022/W33 SP, also made by SKF. Both bearings are the highest precision bearings routinely available from SKF. Bearings having twice this precision are listed in the SKF catalogs but we found that they were not available when we tried ordering them. The cost of the ball bearing system, including materials and manpower costs (at $40/hour) was $8,000. The system is capable of supporting 7 tons. The safety factor is 7.6, which is quite adequate.

The main drawbacks of our ball bearing are that it has too large a coning error (1.5 arcseconds Peak-to-Valley), too much time-varying friction and that it generates vibrations. One can reduce the coning error by lengthening the rod connecting the axial and radial bearings; however it would have to be lengthened to 3-meters for a more acceptable coning error. A problem then arises because a longer rod bends more easily, giving a lower safety coefficient. We measured the torque due to friction, obtaining 1.2 Nm.

B3. Oil bearing.

We bought an oil lubricated bearing but found that it has a safety factor that is much too low for our purpose. The problem comes from the design of this particular oil bearing and is not intrinsic to oil bearings. We measured the torque due to friction, obtaining 6.3 Nm, which is a factor of 5 greater than the one of the ball bearing and 1260 times greater than the one of the air bearing. This is about what we expected since the two bearings have about the same surfaces and the viscosity of the oil that we used is 2500 times the viscosity of air.

B4. Conclusion

In conclusion, our ball bearing is robust, requires little maintenance but has a coning error too large by an order of magnitude. The coefficient of friction is large and varies noticeably in a turn. It also generates a considerable amount of vibrations. Due to its large coning error, it is unsuitable for typical astronomical applications but it may be acceptable for less demanding applications (e.g. as Lidar receiver). It is possible to improve the performance of ball bearing equipped LMs by lengthening the shaft between the trust bearing and the spindle. It should also be possible to make better quality bearings. The oil bearing was inadequate because of its particular design. Better-designed oil bearings may do the job. At the time of this writing, air bearings remain the bearings of choice for astronomical applications. Their main drawbacks are that they are relatively expensive and delicate. However, LMs equipped with air bearings have operated nearly trouble-free for several years in our laboratory and in observatory settings. Their outstanding advantages come from a very low coefficient of friction and the fact that they do not generate vibrations.

C. **Container.**

The container of our 3.7-m, like the design described by Hickson, Gibson & Hogg [17], is made of Kevlar laminated over a foam core. Our experience with the 2.5-m container indicates that analytical computations do not have sufficient accuracy[3]. Our container was therefore designed with finite elements software. The deflections under load measured for the container of the 3.7-meter were found to be in good agreement with the finite element computations. The top surface of the container was spincast with a soft polyurethane resin, having a lower Young modulus than epoxy, that renders the bimetallic plate temperature effect discussed in [13] negligible.

The ball bearing acts as a vibration generator that allows us to study the spectrum of vibrations of the mirror (Fig. 1). For example, we find that the mirror loaded with a 1.85-mm thick layer of mercury has two principal resonance peaks at 12.6 Hz and at 33 Hz. The peak at 12.6 Hz is due to a tilt oscillation of the container. The second peak at 33 Hz is due to a dipolar membrane oscillation of the container: the surface of the container oscillates like the membrane of a drum. The figure is somewhat misleading for it gives the impression that the peak at 33 Hz is more important for the mirror than the peak at 12.6 Hz. In reality this is because the frequency spectrum is measured on the base of the mirror and not on the container. On the container, the tilt oscillation is large and dominates, having amplitude of +- 0.35 arcseconds, which results in a 1.4 arcseconds P-V excursion of the PSF. The amplitude of the 33 Hz oscillation is considerably smaller: nearly a factor of $10^{-2}$ the amplitude of the 12.6 Hz oscillation. As expected from the fact that the mass of mercury is comparable to the mass of the container, we find that the resonant frequencies vary with mercury thickness. The frequency of the tilt oscillation varies between 17 Hz for the empty container to 10.4 Hz for 3 mm of mercury. The frequency of the membrane oscillation varies from 44 Hz for the empty container to 28.8

Hz for 3 mm of mercury. Damping increases with increasing mercury thickness for the tilt oscillation (about a factor of two going from 0.8 to 2-mm thickness) but decreases with increasing thickness for the membrane oscillation (about a factor of two going from 0.8 to 2-mm thickness).

## 2. Instrumentation and Data Analysis

The basic mirror setup and testing facilities are essentially the same as in [2, 3]. We describe them briefly for convenience. We use the Shack cube interferometer [18] mentioned in [3]. The setup of the 3.7-m liquid mirror is similar to the one described in [3]. We use two custom made null lenses designed by C. Morbey that reimage the mirror to f/3.3442. We have thoroughly investigated, with commercial optical design software, the aberrations introduced by alignment errors of the optical components of the null lenses, as well as the effects caused by the coning error of the bearing and the tilt oscillations of the container.

The interferograms are captured with 1/500-second exposure times by a 512X480 CCD detector connected to an 8-bit framegrabber interfaced with a computer. The short integration time is needed because of the oscillation at 12 Hz of the mirror. The interferograms are analyzed with software that uses a Fourier technique [19].

## 3. Wavefronts

The ball bearing generates strong vibrations that induce concentric rings on the mercury surface. Using thin layers of mercury considerably dampens them. However, the surface of our container is parabolic within only 0.3-mm P-V, giving an obvious lower limit to the thickness of the layer. In practice, we found that a thickness of 1.15-mm is about optimal for our particular setup, since thinner layers give a resonant frequency for the tilt oscillation too close to the exciting frequencies of the ball bearing. With this thickness, damping is adequate, except at the center and at the edges of the mirror, where, for technical reasons having to do with thin mercury layers [2], the layer of mercury is deeper. In practice, we could only analyze a surface having a diameter of 3.3-meters, since the outer edges were too perturbed by vibrations induced by the ball bearing

Figure 2 shows a typical interferogram and Figure 3 shows the wavefront obtained from it. The spatial resolution on the mirror is typically 4X8 cm. Table 1 summarizes the results of the analysis of 4 consecutive series of about 50 interferograms (the exact number is given in column 2) each taken over four consecutive rotations of the mirror. The tilts of the individual wavefronts arre removed during data reductions so that the effect of the wobble of the mirror is not included in the statistics. The four series are separated by 5 minutes from each other. Taken at face values the Strehl ratios indicate a mirror that is not diffraction limited but is usable for many astronomical applications. However, the mirror wobbles and the resulting image motion (3.0 arcseconds P-V) considerably worsens the time averaged PSF. The wobble is removed by the data reduction procedures.

To try and determine whether the mirrors had defects that depended on the orientation of the turntable we give in Table 2 average wavefront for 8 azimuth angles. We removed mean values of focus and tilt plus third order coma, spherical aberration, astigmatism. We can see a substantial improvement to the Strehl ratios for they now indicate nearly diffraction limited optics. Removing the aberrations up to third order is justified since simulations with optical design software indicate that the kind of aberrations that we see can easily be generated by misalignments of the null lenses. Experiments and simulations indicate that a lens in our optical setup generates the small amount of astigmatism measured.

Because the mirror is liquid, it can shift shape on time scales of a fraction of a second; hence averaged wavefronts can give a misleadingly optimistic measure of the quality of the mirror. Our past work with liquid mirrors having thin layers of mercury do not indicate that there are substantial time varying phenomena, with the outstanding exception of spiral defects and concentric waves on the mirror. These defects will be discussed in the section on scattered light. Note however that [2] found evidence of traveling waves on a 1.5-m mirror for layer thickness greater than a few mm. These waves were not detected with mercury layers as thin as those used in this work. Table 3 shows the statistics of 3 individual wavefronts captured with the 1/500-second time resolution of our data acquisition system. We see Strehl ratios comparable to those of the average wavefronts in Table 1. Because the mirror wobbles, the alignment between its optical axes and the optical axis of the null lenses varies, inducing a spurious coma. We therefore feel that it is legitimate to remove Coma and defocus. Table 4 shows the statistics of the three wavefronts of Table 3, after removing third order aberrations, indicating a nearly diffraction limited mirror.

Because the mirror spins over a rotating Earth, Coriolis forces are a concern. Quantitative estimates of the effects of the Coriolis forces have been made [20,21], showing that substantial Coma (8 waves P-V) and Astigmatism (22 waves P-V) may be present on a 3.7-m liquid mirror. However, the Coriolis force induces a traveling wave on the mirror and one, intuitively, should expect damping from thin layers and a much reduced amplitude. As an example of a traveling wave dampened by thin layers, see [2]. We obtained wavefronts of a mirror having a 2.3-mm thick mercury layer. It does not show evidence of strong Coriolis-force-induced aberrations. The comatic aberration induced by the Coriolis force can been canceled by the alignment procedure of the null lenses; however this cannot happen for astigmatism. The small value of astigmatism present in our wavefronts is fully compatible with the astigmatism measured on an auxiliary lens and very far from what would be expected from the Coriolis force. Obviously, the effects of the Coriolis force are negligible, presumably because the long-wavelength traveling wave is effectively dampened by the thin mercury layers used.

The main conclusions of the interferometry are:

1. The interferometric measurements are difficult due to the poor quality of the ball bearing that induces ripples on the mirror and an excessive coning error. The coning error is particularly troublesome because our null lens system is very compact and quite sensitive to alignment errors.

2. The most interesting conclusion from our measurements is that the effects of the Coriolis force due to the rotation of the Earth are negligible. This result was anticipated in view of the damping effect of thin mercury layers; it is nonetheless comforting to see it confirmed experimentally. Strong effects due to the Coriolis force were potential "show-stoppers" for liquid mirrors.

3. Considering the poor quality of the bearing, the mirror has a surprisingly good optical quality. The image motion due to the wobbling of the mirror is however too large for the mirror to be useful for astronomical observations. There is little doubt that the surface quality of the mirror would be excellent with a good quality air bearing similar to those used in previous studies.

**4.   Scattered Light**

The interferometry discussed in section 3 has limited spatial resolution (4X8 cm) so that surface defects smaller than the sampling resolution will not be detected. However, such defects introduce scattered light, which can be measured. We have carried out a detailed study of the scattered light generated by the mirror and its causes. The data consist of direct observations of an artificial star created by a laser and a spatial filter. We use the same basic instrumental setup of the interferometry, including the null lenses. Our framegrabber has only 8 bits hence has an insufficient dynamical range to measure an adequate range of the intensity of the PSF, which extends over several orders of magnitude. It is therefore necessary to reconstruct the PSF from series of observations of PSFs having different levels of exposures, some of them with a heavily overexposed core.

Figure 7 in [3] shows a typical sequence of PSFs. That paper discusses the procedure in greater details.

While the previous article [3] describes the main sources of scattered light, and finds that thin mercury layers are very effective at decreasing scattered light, it did not discuss them quantitatively. The ball bearing is a major source of perturbations that induce a considerable amount of scattered light, well above the scattered light seen with a 2.5-m mirror that used an air bearing [3]. We shall make good use of this "feature", for stronger perturbations allow us to better understand and model some of the phenomena (e.g. vibrations) producing scattered light.

Figure 4 gives a very saturated PSF taken with a relatively thick mercury layer (2.3-mm), showing the scattered light within 14 arcseconds of the core of the PSF. We can see a rich structure of scattered light, dominated by conspicuous concentric rings. Fig. 5 shows part of the out-of-focus PSF of the mirror with 2.3 mm of mercury. The out-of-focus PSF is very sensitive to phase error so that Fig. 5 reveals the phase structure of the pupil and the defects that are responsible for the scattered light seen in Fig. 4. Since Fig. 5 is the Fourier transform of Fig. 4, it is not surprising that the most conspicuous defects on the mirror are concentric rings.

The concentric rings are stronger at the center and the edges of the mirror, where they are generated (see section 5) and where the thickness of mercury is greater, for technical reasons [2]: wave damping increases with decreasing thickness [2]. The rings near the edges of the mirror, which is indicated by the arrow, are so strong that we cannot see this region. There are two main reasons for this: First the light is strongly deflected and intercepted by the stops along the optical path. Second, the rings focus light at different locations, above and below the focus of the unperturbed mirror. To highlight the effect of layer thickness on the PSF, we measured the scattered light again after masking a central region having a diameter of 1 meter as well as a 40-cm ring at the edges of the mirror. Losing the central 1-meter actually corresponds to practical observing situations, since instrumentation or a secondary mirror always vignettes the central region of an astronomical mirror. It must be noted, also, that the "full aperture" measurements actually are limited by vignetting in the optical path to a diameter of 3.5-m. Fig. 6 shows encircled energies curves for two mercury thickness with and without mask. The figure shows the large decrease of scattered light that occurs with decreasing mercury thickness, as was noted in previous articles [2,3]. It also shows that the mask has a stronger effect for the 1.15-mm layer than it does for the 2.3-mm layer, confirming that the center and edges are major sources of disturbance. The effect of the mask is stronger with the thinner layer because the thin layer dampens much more effectively the waves over most of its surface, leaving strong waves only at the center and edges, where they are generated and where the mercury layer is thicker.

Although most of this section deals with scattered light generated by vibrations, there are other sources of scattered light, as discussed in [2,3] Turbulence-induced spirals are discussed in section 6. They have relatively high spatial frequency ( of the order of a few cm) and thus are felt in the central part of the PSF ( a few arcseconds). They affect mostly the outer parts of the mirror and are totally dampened out by thin layers. Printthrough-induced scattered light is also present. This phenomenon is discussed in[2] showing that a leveling error of 1 arcsecond introduces a significant printthrough. Since the container wobbles by +- 0.75 arcseconds, we certainly have printthrough-generated scattered light. Thin layers do not improve it.

The main conclusions of the scattered light measurements are:
1. Surface quality increases with decreasing thickness of the mercury layer. This has been known for some time [2,3].
2. For the thinner layers of mercury, the center and the edges are the greatest contributors to scattered light, for the concentric waves are generated in those regions.

3. The positions of the diffraction rings caused by the concentric waves on the mirrors vary with mercury thickness.

**5. Modeling the concentric waves.**

The dispersion relation for waves propagating on a liquid surface is given by [22]

$$\Omega^2 = (gk + \frac{\alpha k^3}{\rho}) \tanh kh, \qquad (1)$$

where g is the acceleration of gravity, k is the wavenumber ($\lambda = \pi/k$), $\rho$ is the density of the liquid, $\alpha$ is the surface tension of the liquid, h the thickness of the liquid layer. The first term in the parenthesis models gravity waves, while the second one refers to capillary waves. In practice, for our setup, the relation between the wavelength of the concentric rings, generated by the resonant frequency of the mirror, and the mercury thickness is more complicated than suggested by Eq. 1, for the resonant frequency of the mirror depends itself on the thickness of mercury. We therefore determined empirically the wavelength-thickness relation by fitting an exponential law to the data. The choice of an exponential law is arbitrary and, for our specific setup and using meter units, gives between thickness of 0.5 and 3.5 mm,

$$\lambda = 0.187 \, h^{0.42}. \qquad (2)$$

It is now straightforward to obtain the relation between the depth of liquid and the position of the rings with the grating law and Eq. 1.

Table 5 summarizes the characteristics of the peaks of the two principal diffraction rings on the surface of the 3.7-m mirror. The amplitudes of other rings are negligible. Wavelengths are obtained from Eq. 1. We find a good agreement between theory and measurements with a surface tension of 0.435 N/m. Figure 7 compares the values of the resonant frequencies of the mirror obtained from the type 2 diffraction rings (points with error bars) to those obtained directly from measurements made with an accelerometer (continuous line). There is good agreement, showing that the type 2 rings are caused by the mirror that vibrates at the resonant frequency. Note that because the positions of the diffraction rings depend on the thickness of the layer and the resonant frequencies of the setup, they can be used to map the thickness of mercury on the surface of the container.

We have identified three types of concentric waves on the surface of the mirror. They are generated at discontinuities in the surface of the liquid (e.g. the circumference and center of the mirror as well as wherever there are holes on the surface of the liquid).

The first type of waves have the largest amplitudes, are generated by the tilt oscillation of the mirror, and oscillate at the resonant frequency of the setup. The pivot point of the tilt oscillation is 70 cm below the surface of the liquid. The tilt oscillation induces a predominantly left to right motion of a cylinder that protrudes at the center of the mirror and a predominantly up and down motion at the circumference of the container. At the edges of the mirror we can assume, with a reasonable approximation, that the waves generated there have amplitudes equal to the motion of the edges. For the 0.7 arcseconds oscillation that we measure, we obtain thus a wave having amplitude of 6 microns P-V at a frequency of 12.1 Hz. At the center, we can assume that the liquid is perturbed by the cylinder that moves mostly with a horizontal periodic motion, a reasonable approximation considering that the pivot is 70 cm below surface. This creates a sinusoidal oscillation of the liquid. The amplitude of the waves can then be computed by using the motion and frequency of the oscillations of the cylinder and normalizing to the volume of the liquid that is displaced. This predicts a maximum slope of 6.4 arcminutes for the wave at the center and 5 arcminutes at the edges, in agreement with the fact that we can see the waves there by looking by eye, on the mirror surface, at the reflection of a sharp edge; as well as by seeing the effect of the waves on the fringes of the interferograms at the center and edges of the mirror.

There are two other types of waves having longer wavelengths and lower frequencies that are present mostly at the center and edges of the mirror. They result from an interaction between the resonant frequency of the setup and the natural resonant frequencies of the deeper mercury layers at the center and edges of the mirror. We do not fully understand the generating mechanism for those waves. They seem to be generated by the sloshing of mercury in the deeper pools of mercury that are present at the center and edges of the mirror, as depicted in fig. 13 of [2]. Although they are not obtained from rigorous analysis, the equations derived below seem to reproduce reasonably well the data.

At the center of the mirror, similar to Fig. 13 of [2], there is a 15-cm diameter pool 2-cm deep. At the center of the pool there is a thin-walled 9 cm diameter aluminum tube that has slots cut in it, to allow the mercury to flow from its inside to the outside. The puddle has a natural oscillation frequency given by

$$n = \frac{1}{2}\sqrt{\frac{g}{pl}} \qquad (3)$$

where $\nu$ is the frequency of oscillation, g the acceleration of gravity and l the diameter of the puddle. Equation 3 predicts $\nu = 2.28$ Hz. However, the puddle does not oscillate at the fundamental frequency, probably due to the presence of the aluminum cylinder. It appears that the puddle oscillates at a higher order frequency, containing 6 nodes and having 2 nodes at the location of the walls of the cylinder. This mode has a frequency of 5.2 Hz, in good agreement with the locations of the peaks of the diffraction rings (type 1 in Table 5). We can find the amplitude of this wave from [23]

$$A(x) = \frac{F_0/m}{|w_0^2 - w^2|}, \qquad (4)$$

All of the terms in Eq. 4 are known, except for $F_0$, which can be obtained from

$$m\frac{d^2x}{dt^2} + kx = F_0 \cos wt, \qquad (5)$$

where the damping term $\kappa x$ is negligibly small. Knowing the frequency and the amplitudes of the oscillations, we find $F_0/m = -(5.8 \times 10^{-3})\cos(76t)$ for 12.1 Hz. This yields an amplitude of 1.2 microns, in agreement with our measurements.

At the rim of the mirror, there is a groove 5-cm wide and 5 mm deep. If we use Eq. 1 for a wave having a wavelength of 10 cm (assuming that the groove contains ½ wave) we obtain a natural frequency of 2.2 Hz, in agreement with the position of a diffraction ring seen with thick mercury layers. These concentric waves can be seen in Fig. 5.

Considering that the amplitude for the centrally-generated or edge-generated waves vary as $1/r$, and considering a damping term $e^{-\alpha r}$, it can be shown [14] that the amplitudes of the waves can be modeled by

$$A(r) = (0.05\, A_c \exp(-\alpha(r-0.05)) + 1.8\, A_e \exp(-\alpha(1.8 - r)))/r, \qquad (6)$$

where $A_c$ and $A_e$ respectively represent the amplitudes at the center and edges. With the expression

$$\zeta(r) = A(r) \cos(2\pi r/\lambda),$$

(7)

we can obtain the amplitudes of the concentric rings on the surface of the mirror for the case where there is constructive interference between the waves originating at the edges and at the center. While this is not necessarily the case and, furthermore, there can be independent oscillations on two axis, this simple hypothesis allows us to gain a quantitative insight into the concentric waves. The quantitative results are also in good agreement with measurements.

We measured the slopes of the concentric waves on the mirror by shining a narrow laser beam on spots at different radii and measuring the deviation of the reflected beam. The measurements are displayed in Fig. 8, along with the amplitudes predicted by Eq. 6 with $\alpha = 3$, obtained from [24]

$$\boldsymbol{a} = \frac{k^3}{\boldsymbol{b}^2 n}\left[\left(\frac{Cosh(4kh)+Cosh(2kh)-1}{Cosh(4kh)-1}\right)+\frac{\boldsymbol{b}}{2kSinh(2kh)}\right]$$

(8)

with

$$n = \frac{1}{2}\left[1+\frac{2kh}{Sinh(2kh)}\right]$$

(9)

and

$$\beta = \sqrt{\frac{\omega}{2\upsilon}},$$

(10)

where $\nu$ is the kinematic viscosity of mercury. The frequency is obtained from Eq. 1 and k by Eq. 2 and $k=2\pi/\lambda$.

We find much better agreement with a larger damping coefficient $\alpha = 7$. Measurements for mercury thickness of 1.8 mm and 1.2 mm are also in better agreement with values of $\alpha$ larger than those predicted by Eq. 8. We find an empirical law for the variation of $\alpha$ with mercury thickness h given by

$$\alpha = 15.8 \, h^{-1.38} \quad . \tag{11}$$

We can now model the concentric waves on the mirror with Eq. 7 and $\alpha$ given by Eq. 11. Figure 9 shows the amplitudes of the waves predicted for a mercury thickness of 2.3-mm, while Fig. 10 shows the amplitudes for a mercury thickness of 0.85-mm, illustrating the spectacular damping properties of thin layers. Figure 11 shows a composite of snapshots of mirror surfaces obtained from out of focus PSFs for varying mercury thickness qualitatively illustrating the damping properties of thin layers. Table 6 gives the RMS surface deviations obtained from the models for varying thickness. We give them for masked as well as for unmasked mirrors (section 4). The table quantifies the dramatic increase in surface quality by going to thinner layers. The large differences between the values for masked and unmasked mirrors show that the waves are strongest at the center and edges, as seen in Fig. 11.

Concentric waves will be considerably weaker for mirrors using air bearings since they are much smoother than our ball bearing. Using the theory of wave generation developed in this section we predict, for typical air bearings such as those used in [2,3], RMS values of $\lambda/51$ with 1.85 mm of mercury, which is negligible.

**6. Rotational-wind induced spirals**

In previous articles we discussed the spiral-shaped defects that are seen on the surfaces of liquid mirrors having thick mercury layers. They can be seen in the right panel of Fig. 12. This is a well-known hydrodynamics phenomenon [25] that has been studied in the laboratory. Comparison between mirrors of different diameters and focal lengths can be made with the help of the Reynold's number

$$R = r^2 \omega/\nu , \qquad (12)$$

where $\nu$ is the kinematic viscosity of air.

Understanding rotational-wind-induced spirals is very important for we suspect that turbulent rotation-induced winds will eventually limit the diameters of liquid mirrors. The values of the Reynold's number at which the spirals appear and the Reynold's number at which they disappear, breaking up in turbulent eddies, are relevant to our discussion.

Experiments with solid rotating disks are summarized by Kobayashi et al. [25], indicating that they begin for $45,000 < R < 232,000$, depending on the studies and techniques used. The situation is just as unsettled for experiments carried out with liquid mirrors. First, let us notice that experiments done on rotating solid disks and liquid mirrors differ somewhat. Liquid mirrors have liquid surfaces, curved surfaces and values of effective gravities that increase with radius. Furthermore, LMs have raised rims, needed to contain the liquid, that interfere with the free flow of air at the edges.

By visually looking at the surfaces of liquid mirrors of varying diameters, one has the impression that the images of the pupils scale with diameters. In other words, the spirals at a radius of 50 cm on a 1.5-m LM look like those at 80 cm on a 2.5-m diameter LM. This would suggest that the spirals are mostly an edge effect. The situation is however rendered more complex by the fact that the larger mirrors are observed with a lower resolution, since the same CCD camera was used. Furthermore, strong concentric waves on the 3.7-m mirror make it even more difficult to see what happens near the center and the edges. Table 7 summarizes the situation for the appearance of the spirals

Prima facie, the Reynold's number marking the appearance of turbulence may seem more relevant to our discussion since, intuitively, one may expect the onset of turbulence to limit the diameters of LMs. Kobayishi et al. [25] quote 265,000 < R < 350,000 with an average of 300,000 for the onset of turbulence. This translates, for the 3.7-m mirror, into 3.8 meters < diameter < 4.4 meters, with an average of 4 meters. Looking at the pupil of the 3.7-m covered with thick layers, we see that the spirals disappear at 1.35 meters < radius <1.5 meters, corresponding to 130,000 < R < 160,000. This is almost a factor of two less than what are measured on solid disks. Beyond the aforementioned radii, we have the strong impression that the spirals break into a speckled pattern. Unfortunately, with the thick layers needed to see the spirals, concentric waves are very strong at those radii, adding noise to the signal that we seek. Using thinner layers does not help detecting the onset of turbulence, since it decreases both the concentric waves and the spirals.

Our experiments seem thus to indicate that the onset of turbulence occurs at lower Reynold's numbers than for rotating disks. This may be due to the fact that our surface is curved or perhaps to the raised rim. On the other hand, it is also clear that using thin layers totally dampen out the defects introduced by turbulence. This is shown in Fig. 12, where we can compare the out of focus PSFs for a mirror with 0.85 mm of mercury and a mirror having 2.3 mm of mercury. The low amplitude defects visible with the 0.85-mm layer are due to a combination of printthrough and seeing cells.

The strong damping seen with thin mercury layers bodes well for LMs in the 8-m class, the next logical step for the technology.

## 7. Conclusion

We have built and tested a 3.7-m diameter liquid mirror. This mirror rotates on a ball bearing, while previous liquid mirrors used air bearings. The ball bearing is less accurate than the air bearings and generates strong vibrations. We discuss briefly the respective advantages and disadvantages of different types of bearings. We find that although the ball bearing has a poor quality, the optical quality of the mirror, with mercury layers 1-mm thick, is surprisingly good. Taken at face values, the instantaneous Strehl ratios indicate a mirror that is not quite diffraction limited but usable for astronomical applications. However, the large coning error of the bearing (1.5 arcseconds P-V) induces an excessive wobble, considerably worsening the time-averaged PSF. The coning error of our ball bearing is definitely not good enough for astronomical observations, but a better designed ball bearing may do the job. There is little doubt that the surface quality of the mirror would be excellent with an air bearing similar to those previously used to make liquid mirrors.

The most interesting result of the interferometry is that we do not see any evidence of the strong astigmatism that may have been expected from Coriolis forces. This is probably because of the strong damping effect of the thin mercury layers.

We have carried out scattered light measurements to detect defects having dimensions smaller than the sampling resolution of the interferometry (4X8cm). We have thus studied two of the major high spatial frequency defects seen on the surface of liquid mirrors: vibration-induced ripples and rotational-wind-induced spirals.

We have made good use of the poor quality of the ball bearing to study the effects of vibrations on the surface of the mirror. We have developed a model that reproduces reasonably well the vibration-induced concentric rings seen on LMs. Our studies of the spirals defects detect the onset of turbulence at smaller Reynold's numbers than we expected. We do not detect neither the spiral defects nor the effects of turbulence for mercury layers 1-mm thick.

This work illustrates, once more the crucial importance of working with mercury layers as thin as possible. Most defects disappear below detection for mercury layers 1-mm thick. Thin layers are crucial for liquid mirrors. Large liquid mirrors would have unacceptable optical qualities for mercury layer much thicker than 1 mm.

On the basis of our tests of the 3.7-m mirror it appears that the upper limits to the diameters of LMs having good bearings are above 4 meters. We make this prediction because the two principal effects that are expected to limit the diameters of LMs, Coriolis forces and self-induced winds, are essentially eliminated by the dampening effects of thin mercury layers.

**Table1.**

RMS wavefront deviations (632.8 Nanometer units), Strehl ratios and third order aberrations (Seidel coefficients) of 4 series of measurements obtained with a mercury thickness of 1,15+/-0,03 mm. The units are wavelength (**632.8 Nanometers**) for the aberration coefficients and degrees for the angles.

| Series | Nber | Tilt | Angle | Focus | Astig | Angle | Coma | Angle | SA3 | Strehl | RMS | P-V |
|---|---|---|---|---|---|---|---|---|---|---|---|---|
| 1 | 51 | 1.525 | 224.2 | 0.630 | 0.445 | -32.8 | 1.997 | 23.3 | -0.512 | 0.12 | 0.274 | 1.489 |
| 2 | 56 | 1.307 | 205.6 | 1.975 | 0.387 | -54.6 | 2.317 | 21.7 | -0.737 | 0.10 | 0.428 | 2.479 |
| 3 | 64 | 0.890 | 213.7 | 0.565 | 0.240 | -18.7 | 1.740 | 18.2 | 0.152 | 0.20 | 0.291 | 1.820 |
| 4 | 47 | 1.030 | -194.6 | 3.032 | 0.563 | -55.2 | 1.895 | 28.5 | -1.598 | 0.05 | 0.476 | 2.513 |
| Mean | 218 | 1.095 | 205.4 | 1.238 | 0.406 | -46.0 | 1.908 | 22.8 | -0.459 | 0.13 | 0.327 | 1.934 |

**Table 2**

Average wavefronts for 8 azimuth angles. We removed mean values of focus and tilt plus third order coma, spherical aberration, and astigmatism. The units are wavelength (**632.8** Nanometers) for the aberration coefficients and degrees for the angles.

|     | Tilt  | Angle   | Focus  | Astig  | Angle | Coma  | Angle | SA3    | Strehl | RMS   | PV    |
|-----|-------|---------|--------|--------|-------|-------|-------|--------|--------|-------|-------|
| 0   | 0.162 | 205.4   | 0.038  | -0.273 | 10.2  | 0.164 | 69.7  | 0.196  | 0.51   | 0.134 | 0.717 |
| 45  | 0.887 | 64.2    | 0.734  | 0.286  | 78.7  | 0.829 | 222.3 | -0.775 | 0.43   | 0.153 | 1.110 |
| 90  | 0.415 | 33.5    | 1.271  | 0.287  | 36.1  | 0.230 | 233.3 | -1.328 | 0.57   | 0.129 | 1.134 |
| 135 | 0.145 | 15.3    | -0.293 | -0.106 | 77.7  | 0.472 | 184.7 | 0.268  | 0.59   | 0.123 | 0.847 |
| 180 | 0.089 | -264.6  | -0.422 | -0.182 | 28.9  | 0.407 | 194.0 | 0.422  | 0.58   | 0.126 | 1.017 |
| 225 | 0.107 | -183.0  | -1.038 | -0.225 | -51.8 | 0.549 | 13.3  | 0.984  | 0.56   | 0.131 | 1.058 |
| 270 | 0.793 | 253.7   | -1.135 | -0.371 | 87.2  | 0.520 | 28.5  | 1.103  | 0.41   | 0.150 | 0.847 |
| 315 | 0.580 | 222.9   | 0.123  | -0.288 | 41.1  | 0.820 | 37.7  | 0.139  | 0.30   | 0.171 | 0.924 |

**Table 3**

Characteristics of 3 individual wavefronts. The units are wavelength (**632.8** Nanometers) for the aberration coefficients and degrees for the angles.

|   | Tilt | Angle | Focus | Astig | Angle | Coma | Angle | SA3 | Strehl | RMS | P-V |
|---|------|-------|-------|-------|-------|------|-------|-----|--------|-----|-----|
| 1 | 3.037 | -188.1 | 1.179 | 0.333 | -31.7 | 3.069 | 10.8 | -1.177 | 0.09 | 0.454 | 2.972 |
| 2 | 1.55 | 214.8 | -0.387 | 0.786 | -36.5 | 1.536 | 48.6 | -1.018 | 0.10 | 0.319 | 2.197 |
| 3 | 0.35 | 223.3 | -0.73 | 0.306 | -35.7 | 1.073 | 26.5 | -0.578 | 0.18 | 0.318 | 2.485 |

**Table 4**

Statistics of the wavefronts of Table 3 after removing tird order aberrations.

|   | RMS ($\lambda$) | | P-V ($\lambda$) | | Strehl | |
|---|---|---|---|---|---|---|
|   | (100%) | (90%) | (100%) | (90%) | (100%) | (90%) |
| 1 | 0.210 | 0.160 | 2.237 | 1.004 | 0.30 | 0.36 |
| 2 | 0.184 | 0.146 | 1.560 | 1.043 | 0.35 | 0.46 |
| 3 | 0.187 | 0.143 | 2.180 | 0.956 | 0.35 | 0.45 |

**Table 5**

Parameters of the principal diffraction rings

| Layer (mm) (+/-0,03) | Type 1 | | | Type 2 | | |
|---|---|---|---|---|---|---|
| | Ring Radius (arcsec) (+/-0,5) | Length (m) (+/- 10%) | Frequency (Hz) | Ring Radius (arcsec) (+/-0,5) | Length (m) +/- 5% | Frequency (Hz) |
| 0.85 | 7.00 | 0.0186 | 5.7+/-0.9 | 14.44 | 0.0090 | 15.5+/-1.4 |
| 1.15 | 5.51 | 0.0236 | 4.9+/-0.6 | 12.16 | 0.0107 | 13.5+/-1.1 |
| 1.85 | 4.29 | 0.0304 | 4.6+/-0.5 | 10.25 | 0.0127 | 12.7+/-0.8 |
| 2.30 | 4.10 | 0.0318 | 4.9+/-0.5 | 8.11 | 0.0161 | 10.2+/-0.6 |

30**Table 6**

RMS surface deviations, as a function of layer thickness, due to vibration-induced concentric rings. They are obtained from the model in section 5.

| Thickness | RMS | | | |
|---|---|---|---|---|
| | Full aperture | | Masked | |
| (mm) | (m) | ($\lambda$) | (m) | ($\lambda$) |
| 0.85 | $4.59 \times 10^{-7}$ | 0.72 | $2.33 \times 10^{-10}$ | 0.00037 |
| 1.15 | $3.50 \times 10^{-7}$ | 0.55 | $3.03 \times 10^{-9}$ | 0.0048 |
| 1.85 | $6.25 \times 10^{-7}$ | 0.99 | $5.68 \times 10^{-8}$ | 0.090 |
| 2.30 | $7.48 \times 10^{-7}$ | 1.18 | $1.57 \times 10^{-7}$ | 0.25 |

**Table 7.**
Diameters and Reynold's numbers at the onset of turbulence as function of mercury thickness

| Reference | Diameter of mirror (m) | R | Hg thickness (mm) |
|---|---|---|---|
| 15 | 1.5 | 9,500 | 5.5 |
| 11 | 1.4 | 9,500 | 1.7 |
| 24 | 2.5 | 37,400 [1] | 2 |
| This work | 3.7 | 40,000 [2] | 2.3 |

Notes

1 Lower limit imposed by optical speckle noise

2 Lower limit imposed by concentric waves

**FIGURE CAPTIONS.**

Figure 1. Vibration spectrum of the mirror. The amplitudes of the oscillations have been measured with an accelerometer. The figure is somewhat misleading for it gives the impression that the peak at 33 Hz is more important for the mirror than the peak at 12.6 Hz. This is not the case (see text for complete explanation).

Figure 2. It shows a typical interferogram of the 3.7-m mirror. The two conspicuous dark rings near the edges are on one of the surfaces of the shack cube and not on the liquid mirror.

Figure 3. It shows the wavefront obtained from the interferogram of Fig. 2. The spatial resolution on the mirror is 4X8 cm.

Figure 4. It gives a very saturated PSF taken with a relatively thick mercury layer (2.3-mm), showing the scattered light within 14 arcseconds of the core of the PSF.

Figure 5. It shows part of the out-of-focus PSF of the mirror with 2.3 mm of mercury. The white arrow indicates the outer edge of the mirror. The black arrow points to a bar that encompasses four 7-cm waves.

Figure 6. It shows encircled energy curves for two mercury thickness with and without mask.

Figure 7. It compares the values of the resonant frequencies of the mirror obtained from the type 2 diffraction rings (points with error bars) to those obtained directly from measurements made with an accelerometer (continuous line).

Figure 8. Slopes of the concentric waves on the mirror. The measurements are displayed along with the amplitudes predicted by the theory described in section 5.

Figure 9. It shows the amplitudes of the waves, predicted by the model in section 5, for a mercury thickness of 2.3-mm

Figure 10. It shows the amplitudes of the concentric waves, predicted by the model in section 5, for a mercury thickness of 0.85-mm.

Figure 11. It shows a composite of snapshots of mirror surfaces obtained from out of focus PSFs for varying mercury thickness qualitatively confirming the damping properties of thin layers.

Figure 12. The left panel shows the surface of the 3.7-m mirror having a mercury thickness of 0.85 mm. Spiral-shaped defects can be seen in the right panel which shows the surface of the 3.7-m mirror having a mercury thickness of 2.3 mm.

34

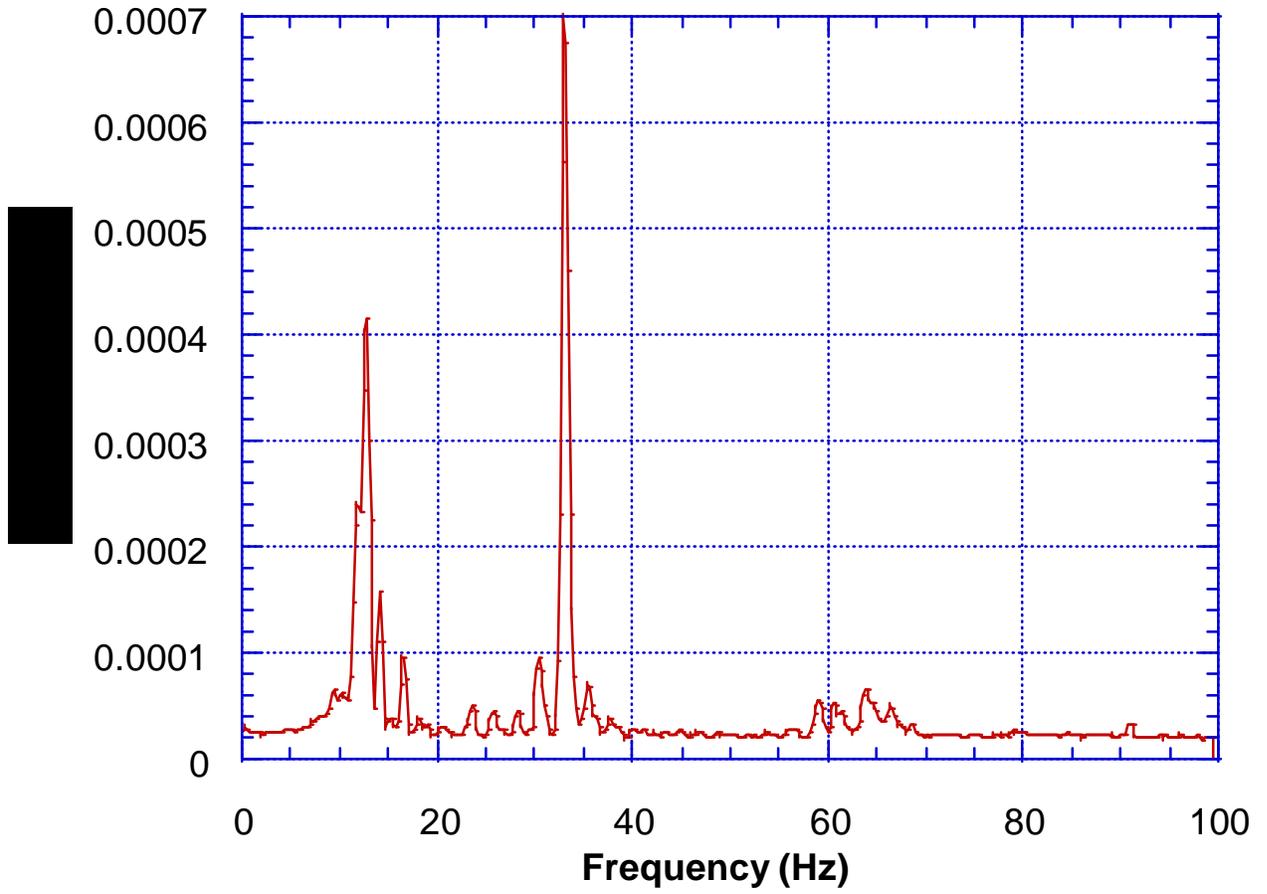

Figure 2

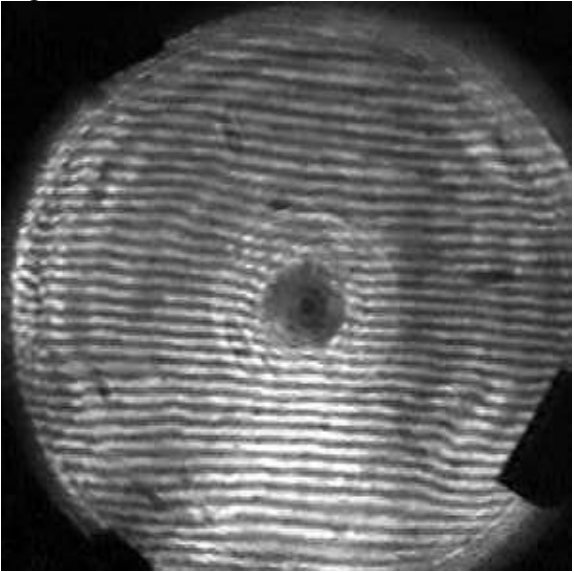

Fig. 3

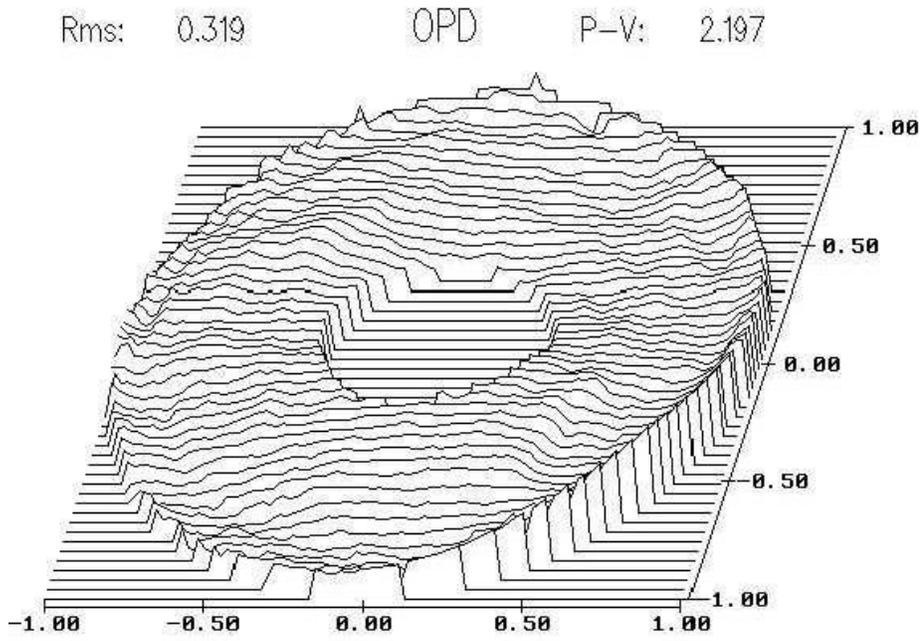

Fig. 4

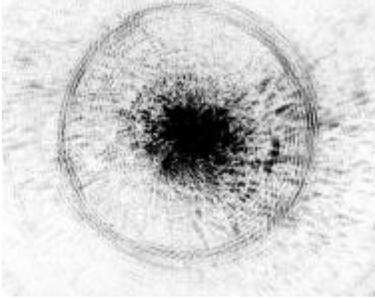

Fig. 5

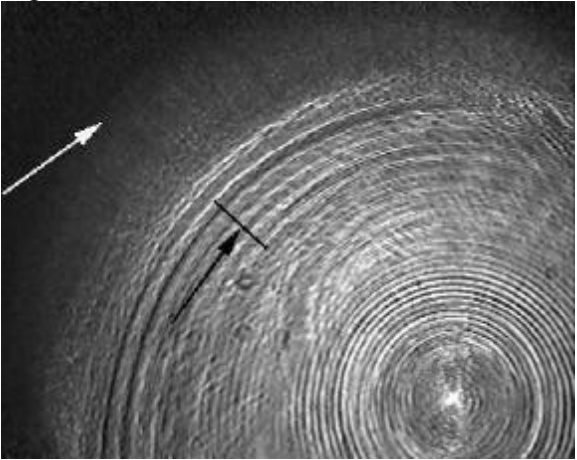

Fig. 6

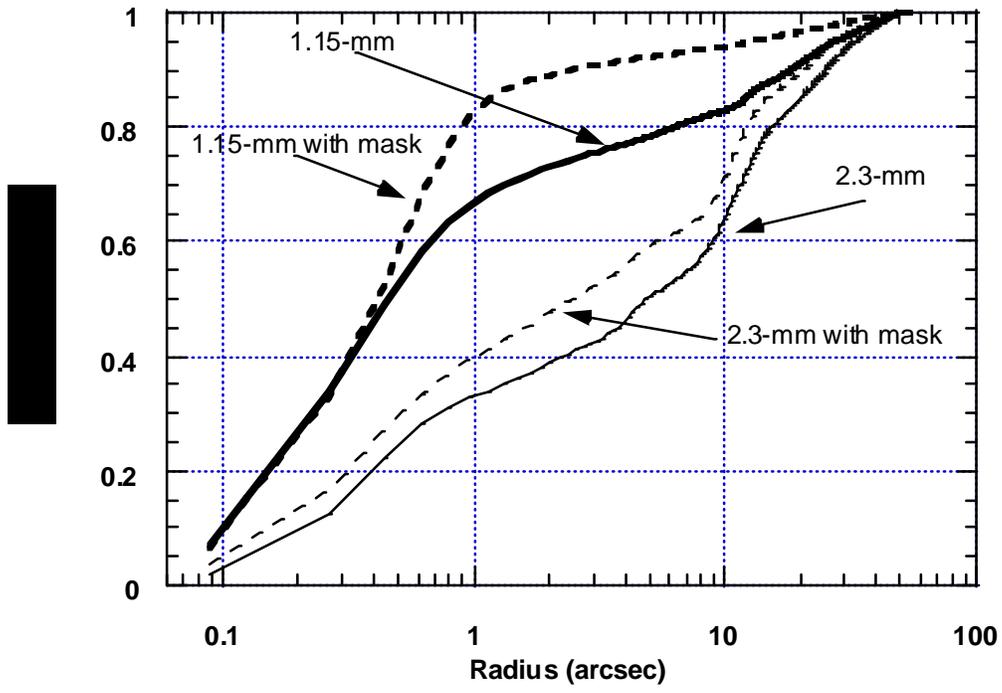

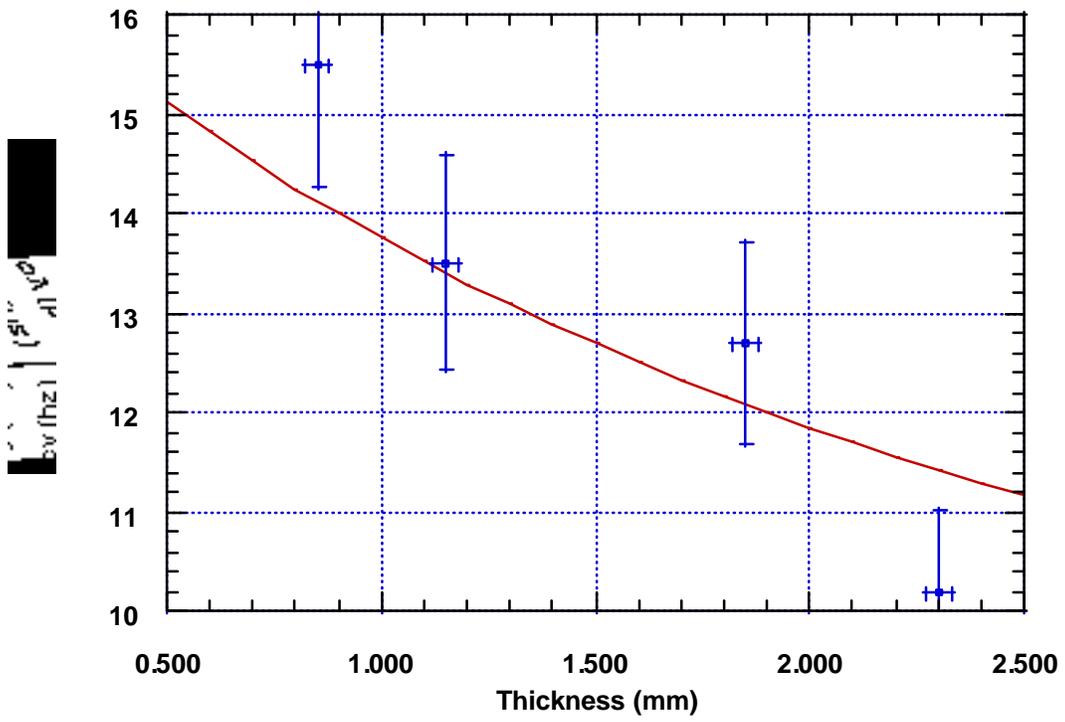

Fig. 7

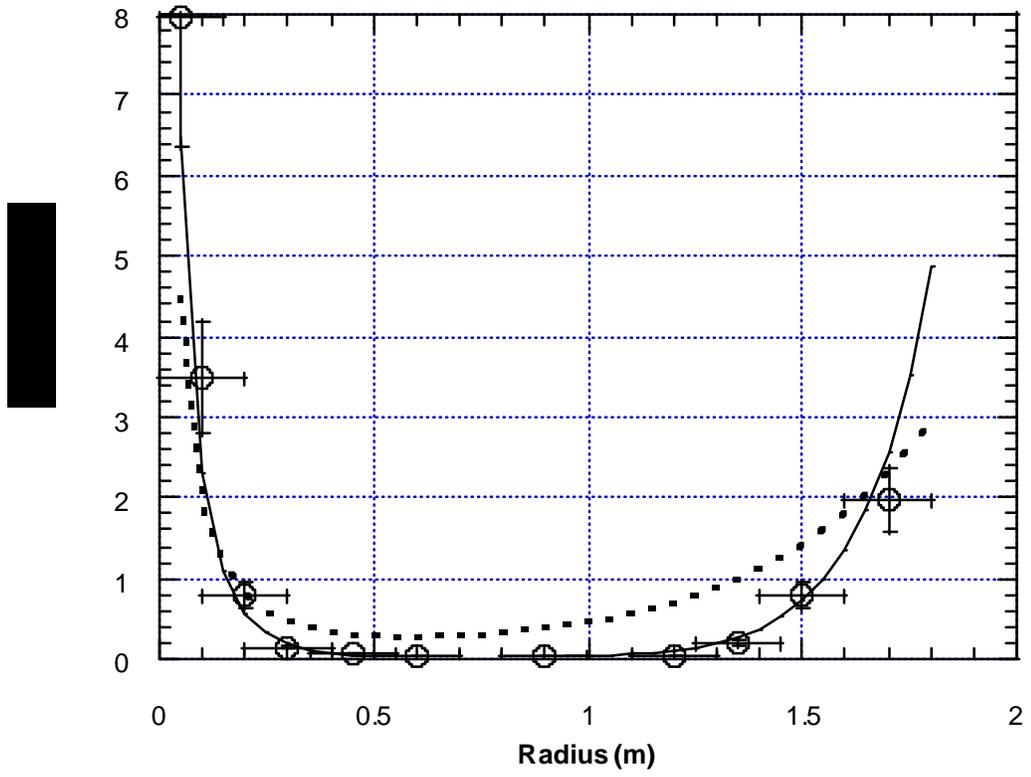

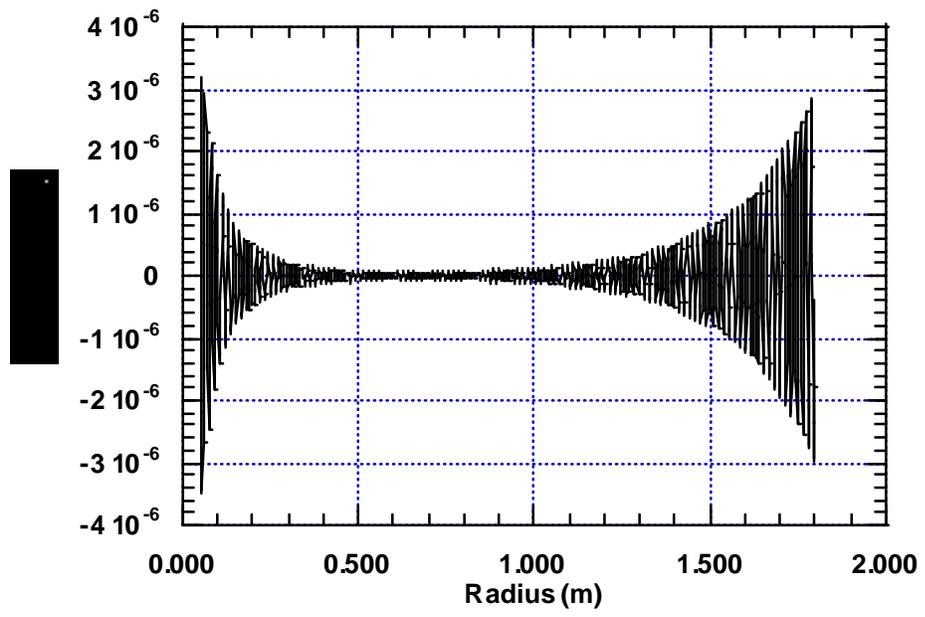

Fig. 9

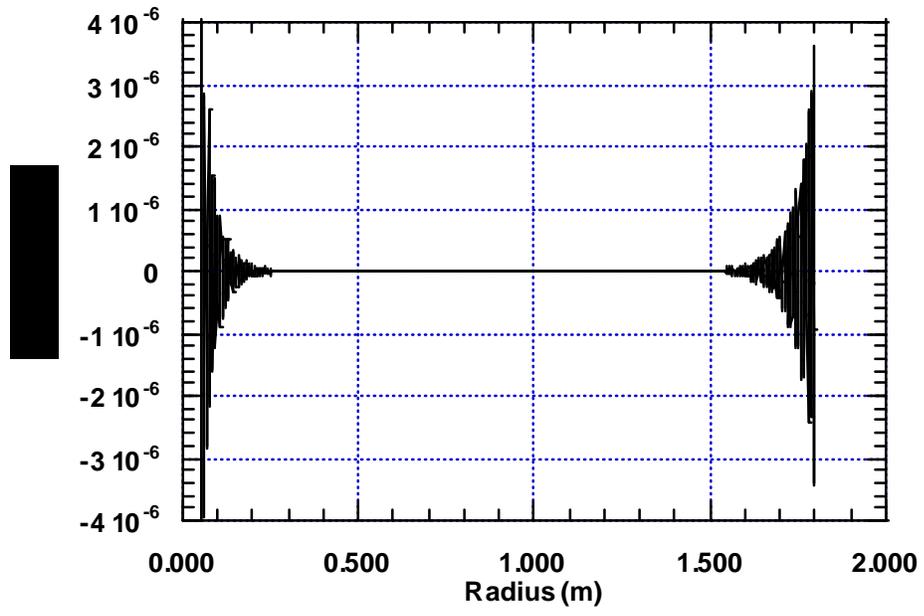

Fig. 10

Fig. 11

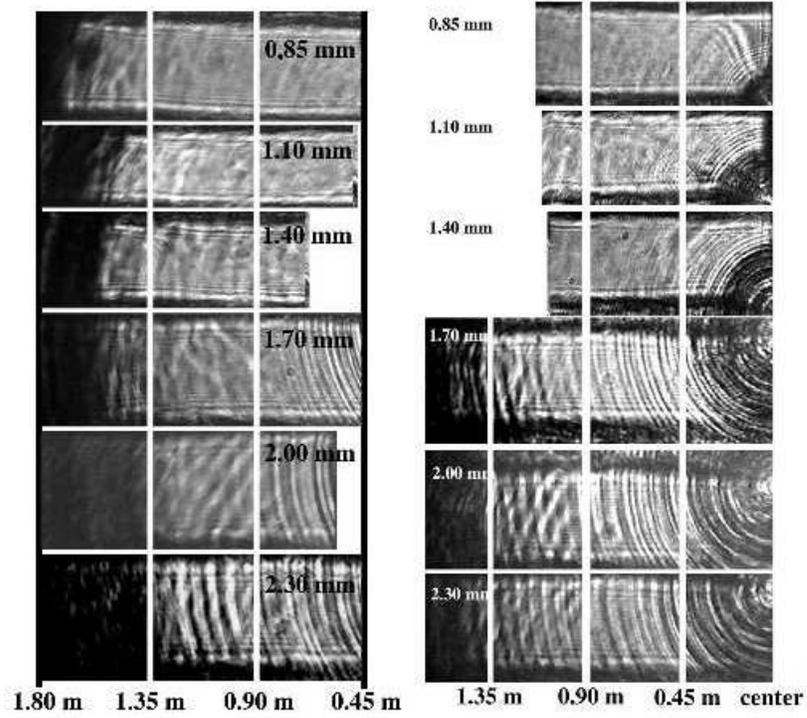

Fig. 12

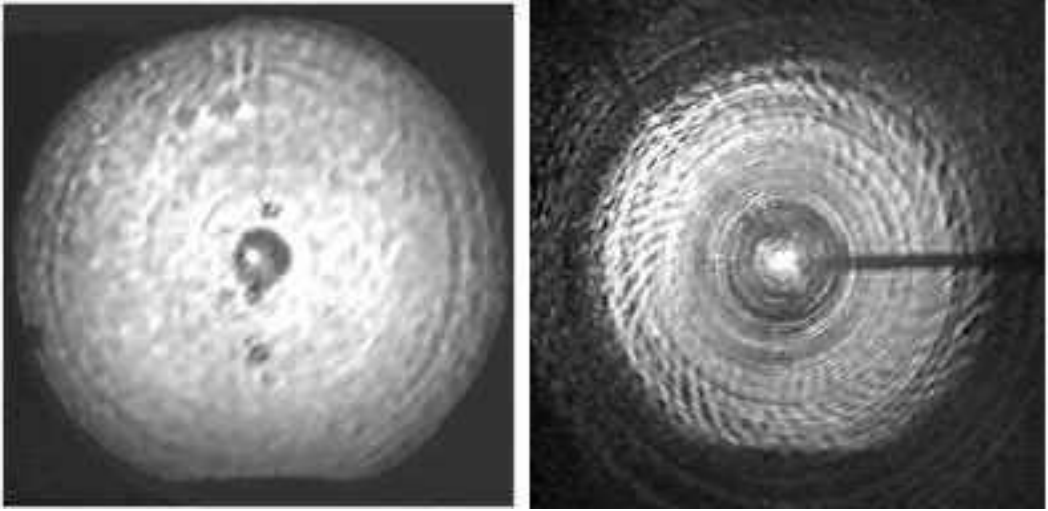